\documentclass[sigconf]{acmart}

\renewcommand\footnotetextcopyrightpermission[1]{} 
\AtBeginDocument{%
  \providecommand\BibTeX{{%
    \normalfont B\kern-0.5em{\scshape i\kern-0.25em b}\kern-0.8em\TeX}}}

\usepackage[nolist,nohyperlinks]{acronym}

\usepackage{mdframed}

\usepackage{enumitem}

\usepackage{subcaption}

\usepackage{csquotes}

\usepackage{pifont} 
\newcommand{\cmark}{\ding{51}}%
\newcommand{\xmark}{\ding{55}}%

\newcommand{\argmax}{\mathrm{arg\,max}}

\usepackage{xspace}
\newcommand{\ie}[0]{\emph{i.e.},\xspace}
\newcommand{\eg}[0]{\emph{e.g.},\xspace}

\begin{document}

\title{Testing the Robustness of Learned Index Structures}

\author{Matthias Bachfischer}
\affiliation{%
  \institution{The University of Melbourne}
  \city{Melbourne}
  \country{Australia}
}
\email{bachfischer.matthias@googlemail.com}

\author{Renata Borovica-Gajic}
\affiliation{%
  \institution{The University of Melbourne}
  \city{Melbourne}
  \country{Australia}}
\email{renata.borovica@unimelb.edu.au}

\author{Benjamin I. P. Rubinstein}
\affiliation{%
  \institution{The University of Melbourne}
  \city{Melbourne}
  \country{Australia}
}
\email{benjamin.rubinstein@unimelb.edu.au}

\renewcommand{\shortauthors}{Bachfischer,  Borovica-Gajic, and Rubinstein}

\begin{abstract}
    While early empirical evidence has supported the case for learned index structures as having favourable average-case performance, little is known about their worst-case performance. By contrast, classical structures are known to achieve optimal worst-case behaviour. 
    This work evaluates the robustness of learned index structures in the presence of adversarial workloads. To simulate adversarial workloads, we carry out a data poisoning attack on linear regression models that manipulates the cumulative distribution function (CDF) on which the learned index model is trained. The attack deteriorates the fit of the underlying ML model by injecting a set of poisoning keys into the training dataset, which leads to an increase in the prediction error of the model and thus deteriorates the overall performance of the learned index structure. 
    We assess the performance of various regression methods and the learned index implementations ALEX and PGM-Index. We show that learned index structures can suffer from a significant performance deterioration of up to 20\% when evaluated on poisoned vs. non-poisoned datasets.
\end{abstract}


\maketitle

\section{Introduction}

In traditional database design, tree-based data structures such as B+ Trees and their variants have seen wide adoption due to their relative ease of implementation and optimal worst-case computational complexity guarantees. B+ Trees are general-purpose data structures that make no prior assumptions about the data distribution and do not take advantage of any patterns that might be specific to the data that the data structure stores. In practice however, real-world data often follows some underlying pattern that, if modeled appropriately, could significantly speed-up the data retrieval process. For instance, given a set of contiguous integer keys (\eg keys from 1 to 100 million), 
the key itself could be used as an offset for the index, thus reducing the time required to look-up any key in the dataset from  $O(\log n)$ for a B+ Tree to $O(1)$. This increase in performance is what Kraska et al. hoped to achieve when they first introduced their work on \ac{LIS} models \cite{RN385}. Even though the concept of a \ac{LIS} is still new, it has already led to a surge of inspiring results that leverage ideas from \ac{ML}, data structures, and database systems \cite{RN388}, \cite{RN417}, \cite{RN408}, \cite{RN455}, \cite{RN416}, \cite{RN453}, \cite{RN414}, \cite{RN407}, \cite{RN457},\cite{RN405}, \cite{RN400}, \cite{RN406}, \cite{RN404}, \cite{RN410}, \cite{RN429}.

The core idea of a \ac{LIS} is to model the functionality of a data structure as a prediction task, \ie given an input key, the objective of the \ac{LIS} is to predict the key's position in a key-sorted collection of key-value pairs. This approach allows the use of continuous functions to encode the data and leverage learning algorithms to approximate key lookup. Specifically, the approach proposed by Kraska et al. \cite{RN385} is to approximate the \ac{CDF} of the keys. Given a key $k$ as an input, the \ac{CDF} returns the probability that the chosen key takes a value less than or equal to $k$ (\ie $P(X \le \text{Key k})$).  
Based on this observation, one can use the \ac{CDF} to compute the number of keys less than the queried key $k$ and infer the key's memory location. In the context of a \ac{LIS}, the \ac{CDF} is used to provide a mapping from the key $k$ to its position in a sorted array. The underlying learning task is one of supervised regression, or simply function interpolation~\cite{RN408}.

While the \ac{LIS} approach may be beneficial in a certain average-case sense, it also carries the risk of exploitation by a malicious adversary. Indeed adversarial analysis is an important tool for understanding worst-case performance and whether \ac{LIS} approaches are worst-case competitive with classical structures.

\section{Related Work}
\subsection{Learned Index Structures}

Since the first published work on learned index structures by Kraska et al. \cite{RN385}, the research community has explored a variety of ideas on how \ac{LIS} could be used as a replacement for traditional index structures such as B+ Trees. Contrary to a B+ Tree, learned index structures rely on an underlying \ac{ML} model to approximate the \ac{CDF}. The most commonly used models are the \acf{PLA} and \acf{LS} models. The \ac{PLA} model is a variant of the linear regression model that tries to approximate the \ac{CDF} of the dataset by dividing it into variable-sized segments. The first and last keys of each segment are then used to construct a linear approximation of the data, resulting in a \ac{PLA} model. As an alternative, \ac{LS} models fit the data by approximating the \ac{CDF} via linear spline points.

\begin{table}[tbh]
  \caption{Overview of \ac{LIS} implementations}
  \vspace{-1em}
  \label{tab:lis-implementations}
  \begin{tabular}{lcccc}
    \toprule
    Index & Model & Updates & Open-Source & Reference\\
    \midrule
      RMI                 & Multiple & \xmark & \cmark & \cite{RN385} \\
      RadixSpline         & LS       & \xmark & \cmark & \cite{RN406} \\
      PGM-Index           & PLA      & \xmark & \cmark & \cite{RN400} \\
      ALEX                & PLA      & \cmark & \cmark & \cite{RN388} \\
      LIPP                & PLA      & \cmark & \cmark & \cite{RN696} \\
      COLIN               & PLA      & \cmark & \xmark & \cite{RN721} \\
  \bottomrule
\end{tabular}
\vspace{-1em}
\end{table}

Table~\ref{tab:lis-implementations} overviews the most relevant learned index implementations published to date. A majority of indexes rely on a single type of model to construct the learned index: \ac{LS} is used by RadixSpline~\cite{RN406} and \ac{PLA} is used by \ac{PGM}-Index~\cite{RN400}, \ac{ALEX}~\cite{RN388} and a variety of other competitors~\cite{RN696}, \cite{RN721}. The \ac{RMI}~\cite{RN385} is the only \ac{LIS} that supports a variety of model types, which gives the \ac{RMI} a greater degree of flexibility, but also increases the complexity of tuning the \ac{RMI} \cite{RN386}.

\subsection{Adversarial Machine Learning}

The term Adversarial Machine Learning describes the study of \acf{ML} techniques against an adversarial opponent that aims to fool a model by supplying deceptive input. Adversarial \ac{ML} has emerged in recent years as a new field, mostly driven by new advancements in computing capabilities \cite{RN701}, \cite{RN727}.  A main domain of focus has been computer vision.

Data poisoning attacks are widely studied within Adversarial \ac{ML} and have been applied in a variety of contexts such as poisoning of neural network or recommender systems \cite{RN447}, \cite{RN705}.  To date, research on data poisoning has mostly been focused on classification and anomaly detection \cite{RN704}, \cite{RN725}, \cite{RN726}, while adversarial regression has largely remained underrepresented \cite{RN459}.

A gradient-based optimization framework for linear classifiers like Lasso or Ridge Regression was first introduced by Xiao et al. \cite{RN706}. Jagielski et al. \cite{RN461} have built upon this work and also proposed a defense mechanism against poisoning attacks called \textit{TRIM}. Another novel attack algorithm on regression learning was proposed by Müller et al. \cite{RN715}. It works by manipulating the training dataset in a way that causes maximum disturbance of the data points. In their experimental evaluation, the authors were able to observe that the \ac{MSE} of the regressor increased by 150 percent after inserting only $2\%$ of poisoned samples.

Poisoning attacks on \ac{LIS} models differ significantly from previous attempts of poisoning linear regression models, because they require poisoning of the \ac{CDF}. This task is challenging, because every insertion affects the values of all points in the dataset. The first researchers to study this new area of research were Kornaropoulos et al., who formulated two poisoning attacks on the hierarchical structure of the \ac{RMI} model \cite{RN415}. By leveraging the attacks described in \cite{RN415}, the researchers were able to increase the error of the poisoned \ac{RMI} model by a factor of up to $300\times$ compared to a non-poisoned model. We expand on this initial idea and perform a comprehensive empirical evaluation across a variety of models (\eg SLR, LogTE, DLogTE, 2P, TheilSen, and LAD discussed in ~\cite{RN696}, as well as ALEX~\cite{RN388} and PGM~\cite{RN400}), open-sourcing our ready-to-use poisoning benchmark to the research community.

\section{Preliminaries}

\subsection{Terminology}

In this work we focus on poisoning attacks against \ac{LIS} models based on \acf{PLA}. To define the model, we let $D=\{x_i,y_i\}_{i=1}^n$ denote the data used by a learned index structure, with $x\in \mathbb{R}$ representing the input data vector and $y\in\mathbb{R}$ representing the output variable. In ordinary linear regression, predictions are made via a linear function $f(x,a,b)=a^{T}x+b$ with parameters $a\in \mathbb{R}^d$ and $b\in \mathbb{R}$  
chosen to minimize average loss  $\mathcal{L}(D,a,b)=\frac{1}{n}\sum_{i=1}^{n}(f(x_i,a,b)-y_i)^2$, also known as the \acl{MSE}. 

In the context of our data poisoning attacks, a key is denoted by $k$ and its key universe (range of potential keys) as $\mathcal{K}$, where $|\mathcal{K}|=m$. Similar to previous work on \ac{LIS} models, it is assumed that keys are given as non-negative integer and that the total order of the keyset can always be derived. It is further assumed that each key is associated with a record and that the records are stored in an in-memory array that is sorted with respect to the key values.

\subsection{Adversarial Model}

\subsubsection{Adversary's Goals}
\label{sec:adversarys_goals} 

When executing a poisoning attack against a \ac{LIS} model, the adversary's goal is to corrupt the learned index model during the training phase (\ie index construction), so that its performance deteriorates during the test phase (\ie prediction of the position of the given key). In this work, we focus on \textit{poisoning availability attacks}, where the adversary's goal is to deteriorate the performance of a learning-based data structure. Specifically, the objective of the adversary is to generate a small number of \textit{poisoning keys} that are used to augment the training dataset that consists of so-called \textit{legitimate keys}. The assumption is that training a \ac{LIS} model with both the \textit{poisoning keys} and \textit{legitimate keys} will result in a model whose performance is worse compared to a \ac{LIS} model that is trained on only the \textit{legitimate keys}.

\subsubsection{Adversary's Capabilities \& Knowledge}
\label{sec:adversarys_capabilities_and_knowledge}
Data poisoning attacks usually distinguish between two attack scenarios: \textit{white-box} and \textit{black-box} poisoning attacks. 

In this work, we focus on white-box poisoning attacks where the attacker is assumed to have full access to the training data, \ie the keyset $K$ and the slope and intercept parameters $a$ and $b$ of the linear regression \cite{RN703}, \cite{RN704}, \cite{RN415}. When performing the attack, the adversary is able to inject up to $p$ maliciously-crafted poisoning keys into the training set prior to training the \ac{LIS} model.
The total number of data points in the training set is given by $N = n + p$, where $n$ denotes the number of legitimate keys and $p$ the number of poisoning keys in the training data. Similar to previous work on poisoning attacks, it is assumed that the adversary is able to control only a tiny fraction of the training set limited by the poisoning percentage $\alpha$, where $\alpha = p / n$  \cite{RN461}, \cite{RN706}.

If we were to adapt this attack to the \textit{black-box} scenario where the adversary has no direct access to the training data or training parameters of the model, the attacker would first need to infer the parameters of the model and subsequently use their estimates to perform the poisoning attack. Though more difficult to execute, black-box attacks allow better transferability of poisoning attacks against different training sets, as shown in \cite{RN706} and \cite{RN705}.

\subsubsection{Attack Evaluation Metric}
\label{sec:attack_evaluation_metric}
In this research, the effect of poisoning the \ac{LIS} is measured with respect to the mean lookup time in nanoseconds. To measure the performance impact of the poisoning attack, we calculate the ratio of the mean lookup time of a model that is trained on the legitimate data and the mean lookup time of a model trained on the poisoned data.
For completeness, we also report the mean lookup time (in nanoseconds) across all poisoning thresholds for all the models considered.

\section{Testing the Robustness of Learned Index Structures}

To test the robustness of learned index structures, we execute a poisoning attack on linear regression models by attacking the \ac{CDF} of the training data. The attack works by inserting a certain number of \textit{poisoning keys} into the training dataset with the aim of increasing the approximation error of the regression and thus deteriorating the overall performance of the index.

To formulate the poisoning attack, we consider a \ac{LIS} to consist of an index that is being constructed on a keyset $K$ of size $n$, where each key $k \in K$ has a rank $r$ in the interval $[1, n]$. Here, $r$ denotes the position of $k$ in an ordered sequence of $K$. The objective of the \ac{LIS} is to approximate the rank of the queried key by constructing a regression model on $(k, r)$, where the X-value is given by the key $k$, and the Y-value is denoted by the rank $r$. In other words, the function that the regression model approximates is the \ac{CDF} of the input dataset.

Prior work on poisoning attacks on linear regression models was aimed at inserting maliciously-crafted poisoning keys that cause a \enquote{local change}, \ie inserting keys that do not affect the X- and Y-values of the legitimate points \cite{RN461}. In the case of \ac{LIS} models, the insertion of a single, maliciously-crafted key $k_p$ will cause a shift in the rank of all keys larger than $k_p$. This change will in turn trigger a shift of the \ac{CDF}, thus compounding the effect of the adversarial insertion.

To this date, the \enquote{compounding effect of adversarial insertion} has only been studied by the authors of \cite{RN415}, where they introduced a novel poisoning attack for linear regression on \acp{CDF} called \textit{GreedyPoisoningRegressionCDF}. We follow their approach and describe the poisoning strategy of an adversary targeting linear regression models on \acp{CDF} in Definition \ref{def:poisoning_linear_regression_on_cdf}. The parameter $\lambda$ denotes the upper bound that limits the size of the poisoning keyset $P$ and is chosen to be proportional to the size of the keyset. In the experiments described in Section ~\ref{sec:experimental_setup}, $\lambda$ was set to a range of values between $\lambda=0.01n$ and $\lambda=0.20n$. For further details on the poisoning algorithm, the interested reader is referred to \cite{RN415}.

\mdfdefinestyle{FrameDefinition}{%
    linecolor=black,
    outerlinewidth=1pt,
    innertopmargin=-2pt,
    innerbottommargin=-2pt,
    innerrightmargin=1pt,
    innerleftmargin=1pt,
        leftmargin = 1pt,
        rightmargin = 1pt
}

\begin{mdframed}[style=FrameDefinition,nobreak=false,align=center]
\begin{definition}[\textbf{Poisoning Linear Regression on CDF}]
    Let $K$ be the set of $n$ integers that correspond to the keys and let $P$ be the set of $p$ integers that comprise the poisoning keys. 
    The augmented set on which the linear regression model is trained is $\{(k'_1, r'_1), (k'_2, r'_2), \cdots, (k'_{n'}, r'_{n'})\}$, where $k'_i\in K\cup P$ and $r'_i \in [1,n+p]$. 
    The goal of the adversary is to choose a set $P$ of size at most $\lambda$ so as to maximize the loss function of the augmented set $K\cup P$:
    \[\argmax_{P\mathrm{~s.t.~}|P|\leq \lambda}\left(\min_{a,b}\mathcal{L}\left(\{k'_i,r'_i\}_{i=1}^{n+p},a,b\right)\right)\]
    \label{def:poisoning_linear_regression_on_cdf}
\end{definition}
\end{mdframed}

\section{Experimental Evaluation}

\subsection{Experimental Setup}
\label{sec:experimental_setup}
To test the robustness of learned index structures, we set-up a flexible microbenchmark that allows us to quickly test the robustness of learned index structures against data poisoning attacks. The microbenchmark is based on the source code that was published by Eppert et al. \cite{RN696}. We have extended the existing microbenchmark by implementing the \ac{CDF} poisoning attack against different types of regression models and the learned index implementations \ac{ALEX} and \ac{PGM}-Index. The corresponding source code used in this work is available online.\footnote{\url{https://github.com/Bachfischer/LogarithmicErrorRegression}}

\begin{figure}[h]
    \centering
      \includegraphics[width=0.48\textwidth]{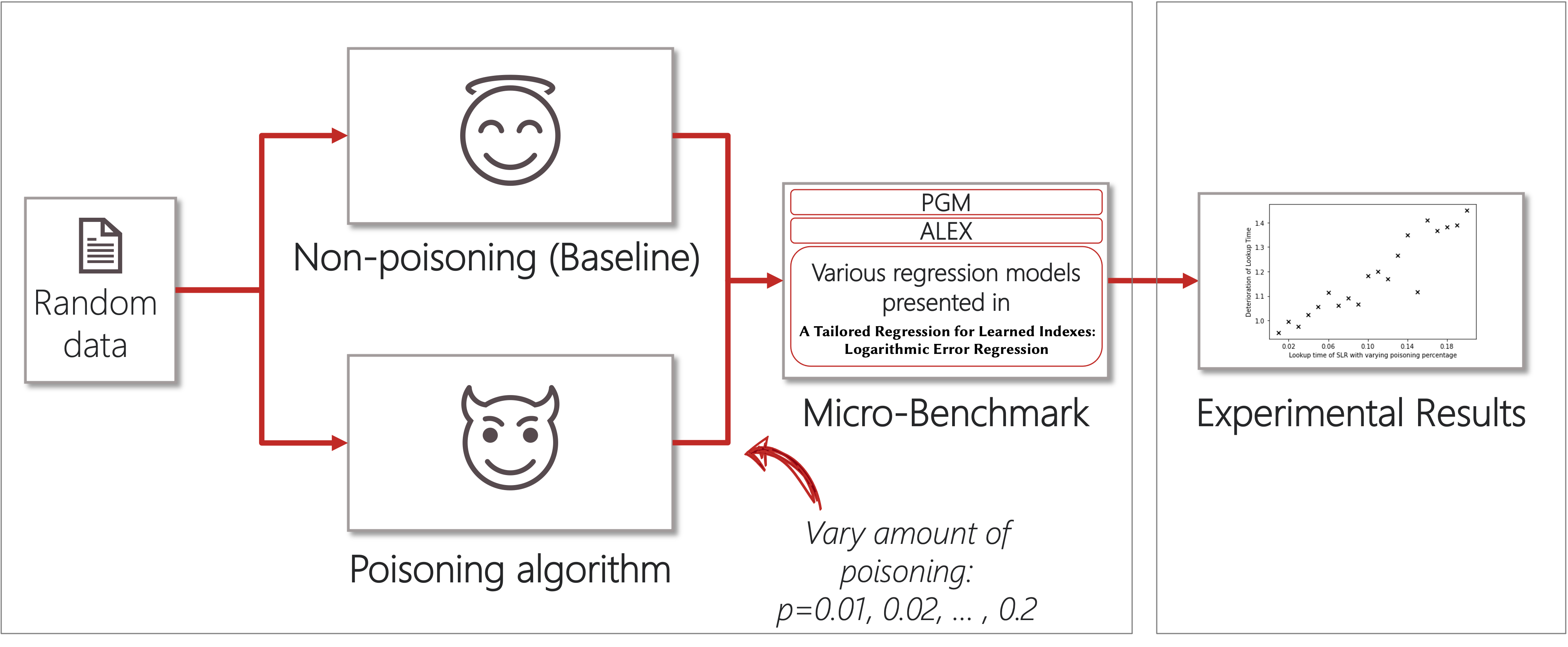}
         \vspace{-0.5em}
        \caption{Benchmarking architecture used for experiments.}
        \vspace{-0.5em}
        \label{fig:system-architecture}
\end{figure}
  
The system architecture that we used for our experiments is shown in Figure~\ref{fig:system-architecture}. To simulate database workload, we have first generated a synthetic dataset consisting of 1000 keys. We subsequently executed the poisoning attack against the synthetic dataset while varying the poisoning threshold parameter from $p=0.01$ to $p=0.20$. For each poisoning threshold, we obtained the corresponding set of poisoning keys and used the legitimate and poisoned keysets to measure the mean lookup time of the indexes on the legitimate (non-poisoned) dataset and the mean lookup time on the poisoned dataset.

\begin{figure*}[h]
    \centering
    \begin{subfigure}[b]{0.24\textwidth}
        \centering
        \includegraphics[width=\textwidth]{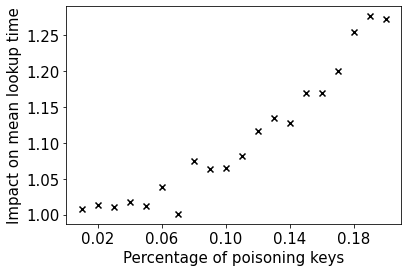}
        \caption{SLR}
    \end{subfigure}
    \hfill
    \begin{subfigure}[b]{0.24\textwidth}
        \centering
        \includegraphics[width=\textwidth]{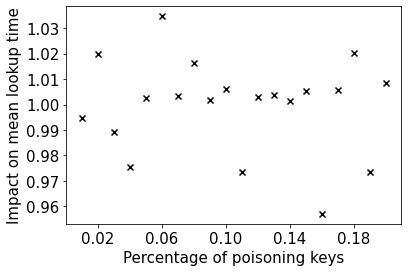}
        \caption{LogTE}
    \end{subfigure}
    \hfill
    \begin{subfigure}[b]{0.24\textwidth}
        \centering
        \includegraphics[width=\textwidth]{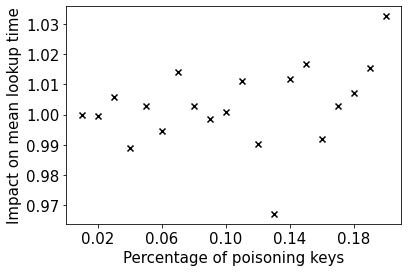}
        \caption{DLogTE}
    \end{subfigure}
    \hfill
    \begin{subfigure}[b]{0.24\textwidth}
        \centering
        \includegraphics[width=\textwidth]{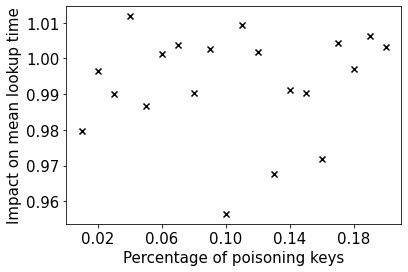}
        \caption{2P}
    \end{subfigure}
    \\
        \begin{subfigure}[b]{0.24\textwidth}
        \centering
        \includegraphics[width=\textwidth]{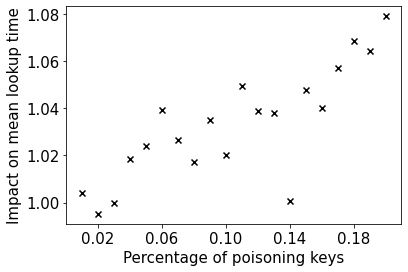}
        \caption{TheilSen}
    \end{subfigure}
    \hfill
    \begin{subfigure}[b]{0.24\textwidth}
        \centering
        \includegraphics[width=\textwidth]{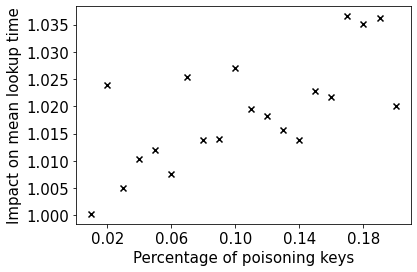}
        \caption{LAD}
    \end{subfigure}
    \hfill
    \begin{subfigure}[b]{0.24\textwidth}
        \centering
        \includegraphics[width=\textwidth]{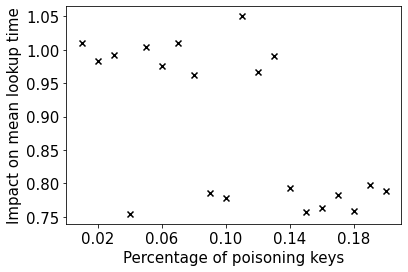}
        \caption{ALEX}
    \end{subfigure}
    \hfill
    \begin{subfigure}[b]{0.24\textwidth}
        \centering
        \includegraphics[width=\textwidth]{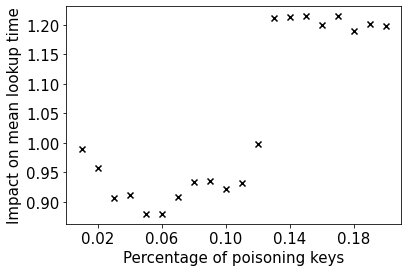}
        \caption{PGM}
    \end{subfigure}
    \caption{Performance deterioration of LIS models under poisoning attacks.}
    \label{fig:performance_deterioration}
    \vspace{-1em}
\end{figure*}

To test the robustness of learned index structures, we focused on \ac{LIS} models that approximate the \ac{CDF} via \ac{PLA} and whose source code is available as open-source. We have therefore decided to include \ac{ALEX}, \ac{PGM}-Index, and the regression models\footnote{Regression models: SLR, LogTE, DLogTE, 2P, TheilSen, and LAD} discussed in \cite{RN696} into our benchmark. Additional details on the indexes are provided in Table~\ref{tab:evaluated_indexes}.

\begin{table}[tbh]
  \begin{tabular}{llll}
        \toprule
        Method & Description & Parameters & Source \\
        \midrule
        Regressions (SLR etc.) & \cite{RN696} & - & \cite{sourcecode-logarithmic-error-regression} \\
        ALEX & \cite{RN388} & - & \cite{sourcecode-alex} \\
        PGM-Index & \cite{RN400} & max. error $\epsilon$ & \cite{sourcecode-pgm} \\
        \bottomrule
    \end{tabular}
  \caption{Overview of evaluated indexes.}
  \label{tab:evaluated_indexes}
\end{table}

\vspace{-2em}
\subsection{Experimental Results}

Figure~\ref{fig:performance_deterioration} shows the results of the experiments. The performance deterioration of the \ac{LIS} is calculated as the ratio between the mean lookup time in nanoseconds for the poisoned datasets and the mean lookup time for the legitimate (non-poisoned) dataset.

From the graphs, we can observe that simple linear regression (SLR) is particularly prone to the poisoning attack, as this regression model shows a steep increase in the mean lookup time when evaluated on the poisoned data. The performance of the competitors that optimize a different error function such as LogTE, DLogTE and 2P are more robust against adversarial attacks. For these regression models, the mean lookup time remains relatively stable even when the poisoning threshold is increased substantially.

Because SLR is the de-facto standard in learned index structures and used internally by the \ac{ALEX} and the \ac{PGM}-Index implementations, we would expect that these two models also exhibit a relatively high performance deterioration when evaluated on the poisoned dataset. Surprisingly, ALEX does not show any significant performance impact. This is most likely due to the usage of gapped arrays that allows the model to easily capture outliers in the data (up to a certain point). The performance of the PGM-Index deteriorates by a factor of up-to 1.3$\times$.

\begin{figure}[h]
    \centering
      \includegraphics[width=0.44\textwidth]{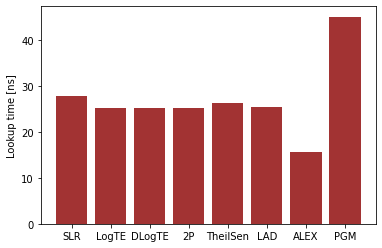}
      \vspace{-0.5em}
        \caption{Mean lookup time  avg.
across all poisoning levels.}
        \vspace{-0.5em}
        \label{fig:lookup}
\end{figure}

For further analysis, we have also calculated the overall mean
lookup time for the evaluated learned indexes, averaged across
all experiments. The results are shown in Figure \ref{fig:lookup}. From Figure \ref{fig:lookup},
we can observe that ALEX dominates all learned index structures.
The performance of the regression models SLR, LogTE, DLogTE,
2P, TheilSen and LAD is also relatively similar, in a range between
30 - 40 nanoseconds, while the PGM-Index performs worst with a mean lookup time of >
50 nanoseconds.

\section{Conclusions and future outlook}

In this research, we have tested the robustness of a variety of regression models as well as two learned index implementations \ac{ALEX} and \ac{PGM}-Index against adversarial workload. To simulate adversarial workload, we have executed a poisoning attack on a synthetic dataset consisting of 1000 keys.
To evaluate the success of the poisoning attack, we measured the performance deterioration of the lookup time for various indexes. The results show that \ac{LIS} models are prone to poisoning attacks and exhibit significantly worse performance after only a small subset of poisoning keys is introduced into the keysets.

The experiments described in this research focused exclusively on poisoning models that try to approximate the \ac{CDF} via linear regression. Recent publications have introduced other models for constructing a \ac{LIS} such as polynomial interpolation \cite{RN408} and logarithmic error regression \cite{RN696}. These novel model architectures would also provide an interesting target for poisoning attacks.
While the poisoning attack used in this paper works by introducing poisoning keys prior to training the index models, future research may investigate how an adversary could leverage the update functionality of dynamic learned index models to insert and remove keys from a trained \ac{LIS} model during runtime to deteriorate the fit of the \ac{LIS}.

\bibliographystyle{ACM-Reference-Format}
\bibliography{bibliography}


\begin{thebibliography}{35}


\ifx \showCODEN    \undefined \def \showCODEN     #1{\unskip}     \fi
\ifx \showDOI      \undefined \def \showDOI       #1{#1}\fi
\ifx \showISBNx    \undefined \def \showISBNx     #1{\unskip}     \fi
\ifx \showISBNxiii \undefined \def \showISBNxiii  #1{\unskip}     \fi
\ifx \showISSN     \undefined \def \showISSN      #1{\unskip}     \fi
\ifx \showLCCN     \undefined \def \showLCCN      #1{\unskip}     \fi
\ifx \shownote     \undefined \def \shownote      #1{#1}          \fi
\ifx \showarticletitle \undefined \def \showarticletitle #1{#1}   \fi
\ifx \showURL      \undefined \def \showURL       {\relax}        \fi
\providecommand\bibfield[2]{#2}
\providecommand\bibinfo[2]{#2}
\providecommand\natexlab[1]{#1}
\providecommand\showeprint[2][]{arXiv:#2}

\bibitem[Abu-Libdeh et~al\mbox{.}(2020)]%
        {RN416}
\bibfield{author}{\bibinfo{person}{Hussam Abu-Libdeh}, \bibinfo{person}{Deniz
  Altınbüken}, \bibinfo{person}{Alex Beutel}, \bibinfo{person}{Ed~H Chi},
  \bibinfo{person}{Lyric Doshi}, \bibinfo{person}{Tim Kraska},
  \bibinfo{person}{Andy Ly}, {and} \bibinfo{person}{Christopher Olston}.}
  \bibinfo{year}{2020}\natexlab{}.
\newblock \showarticletitle{Learned Indexes for a Google-scale Disk-based
  Database}.
\newblock \bibinfo{journal}{\emph{arXiv preprint arXiv:2012.12501}}
  (\bibinfo{year}{2020}).
\newblock


\bibitem[Barreno et~al\mbox{.}(2010)]%
        {RN727}
\bibfield{author}{\bibinfo{person}{Marco Barreno}, \bibinfo{person}{Blaine
  Nelson}, \bibinfo{person}{Anthony~D Joseph}, {and} \bibinfo{person}{J~Doug
  Tygar}.} \bibinfo{year}{2010}\natexlab{}.
\newblock \showarticletitle{The security of machine learning}.
\newblock \bibinfo{journal}{\emph{Machine Learning}} \bibinfo{volume}{81},
  \bibinfo{number}{2} (\bibinfo{year}{2010}), \bibinfo{pages}{121--148}.
\newblock
\showISSN{1573-0565}


\bibitem[Biggio et~al\mbox{.}(2012)]%
        {RN704}
\bibfield{author}{\bibinfo{person}{Battista Biggio}, \bibinfo{person}{Blaine
  Nelson}, {and} \bibinfo{person}{Pavel Laskov}.}
  \bibinfo{year}{2012}\natexlab{}.
\newblock \showarticletitle{Poisoning attacks against support vector machines}.
\newblock \bibinfo{journal}{\emph{arXiv preprint arXiv:1206.6389}}
  (\bibinfo{year}{2012}).
\newblock


\bibitem[Biggio and Roli(2018)]%
        {RN726}
\bibfield{author}{\bibinfo{person}{Battista Biggio} {and}
  \bibinfo{person}{Fabio Roli}.} \bibinfo{year}{2018}\natexlab{}.
\newblock \showarticletitle{Wild patterns: Ten years after the rise of
  adversarial machine learning}.
\newblock \bibinfo{journal}{\emph{Pattern Recognition}}  \bibinfo{volume}{84}
  (\bibinfo{year}{2018}), \bibinfo{pages}{317--331}.
\newblock
\showISSN{0031-3203}


\bibitem[Boffa et~al\mbox{.}(2021)]%
        {RN414}
\bibfield{author}{\bibinfo{person}{Antonio Boffa}, \bibinfo{person}{Paolo
  Ferragina}, {and} \bibinfo{person}{Giorgio Vinciguerra}.}
  \bibinfo{year}{2021}\natexlab{}.
\newblock \showarticletitle{A “Learned” Approach to Quicken and Compress
  Rank/Select Dictionaries}. In \bibinfo{booktitle}{\emph{2021 Proceedings of
  the Workshop on Algorithm Engineering and Experiments (ALENEX)}}.
  \bibinfo{publisher}{SIAM}, \bibinfo{pages}{46--59}.
\newblock


\bibitem[Dai et~al\mbox{.}(2020)]%
        {RN417}
\bibfield{author}{\bibinfo{person}{Yifan Dai}, \bibinfo{person}{Yien Xu},
  \bibinfo{person}{Aishwarya Ganesan}, \bibinfo{person}{Ramnatthan Alagappan},
  \bibinfo{person}{Brian Kroth}, \bibinfo{person}{Andrea Arpaci-Dusseau}, {and}
  \bibinfo{person}{Remzi Arpaci-Dusseau}.} \bibinfo{year}{2020}\natexlab{}.
\newblock \showarticletitle{From wisckey to bourbon: A learned index for
  log-structured merge trees}. In \bibinfo{booktitle}{\emph{14th {USENIX}
  Symposium on Operating Systems Design and Implementation ({OSDI} 20)}}.
  \bibinfo{pages}{155--171}.
\newblock
\showISBNx{193913319X}


\bibitem[Ding et~al\mbox{.}(2020a)]%
        {RN388}
\bibfield{author}{\bibinfo{person}{Jialin Ding}, \bibinfo{person}{Umar~Farooq
  Minhas}, \bibinfo{person}{Jia Yu}, \bibinfo{person}{Chi Wang},
  \bibinfo{person}{Jaeyoung Do}, \bibinfo{person}{Yinan Li},
  \bibinfo{person}{Hantian Zhang}, \bibinfo{person}{Badrish Chandramouli},
  \bibinfo{person}{Johannes Gehrke}, {and} \bibinfo{person}{Donald Kossmann}.}
  \bibinfo{year}{2020}\natexlab{a}.
\newblock \showarticletitle{ALEX: an updatable adaptive learned index}. In
  \bibinfo{booktitle}{\emph{Proceedings of the 2020 ACM SIGMOD International
  Conference on Management of Data}}. \bibinfo{pages}{969--984}.
\newblock


\bibitem[Ding et~al\mbox{.}(2020b)]%
        {RN404}
\bibfield{author}{\bibinfo{person}{Jialin Ding}, \bibinfo{person}{Vikram
  Nathan}, \bibinfo{person}{Mohammad Alizadeh}, {and} \bibinfo{person}{Tim
  Kraska}.} \bibinfo{year}{2020}\natexlab{b}.
\newblock \showarticletitle{Tsunami: A learned multi-dimensional index for
  correlated data and skewed workloads}.
\newblock \bibinfo{journal}{\emph{arXiv preprint arXiv:2006.13282}}
  (\bibinfo{year}{2020}).
\newblock


\bibitem[Eppert et~al\mbox{.}(2021a)]%
        {RN696}
\bibfield{author}{\bibinfo{person}{Martin Eppert}, \bibinfo{person}{Philipp
  Fent}, {and} \bibinfo{person}{Thomas Neumann}.}
  \bibinfo{year}{2021}\natexlab{a}.
\newblock \showarticletitle{A Tailored Regression for Learned Indexes:
  Logarithmic Error Regression}.
\newblock  (\bibinfo{year}{2021}).
\newblock


\bibitem[Eppert et~al\mbox{.}(2021b)]%
        {sourcecode-logarithmic-error-regression}
\bibfield{author}{\bibinfo{person}{Martin Eppert}, \bibinfo{person}{Philipp
  Fent}, {and} \bibinfo{person}{Thomas Neumann}.}
  \bibinfo{year}{2021}\natexlab{b}.
\newblock \bibinfo{booktitle}{\emph{A Tailored Regression for Learned Indexes:
  Logarithmic Error Regression}}.
\newblock
\urldef\tempurl%
\url{https://github.com/umatin/LogarithmicErrorRegression}
\showURL{%
\tempurl}


\bibitem[Ferragina and Vinciguerra(2020)]%
        {RN400}
\bibfield{author}{\bibinfo{person}{Paolo Ferragina} {and}
  \bibinfo{person}{Giorgio Vinciguerra}.} \bibinfo{year}{2020}\natexlab{}.
\newblock \showarticletitle{The PGM-index: a fully-dynamic compressed learned
  index with provable worst-case bounds}.
\newblock \bibinfo{journal}{\emph{Proceedings of the VLDB Endowment}}
  \bibinfo{volume}{13}, \bibinfo{number}{10} (\bibinfo{year}{2020}),
  \bibinfo{pages}{1162--1175}.
\newblock
\showISSN{2150-8097}


\bibitem[Ferragina and Vinciguerra(2021)]%
        {sourcecode-pgm}
\bibfield{author}{\bibinfo{person}{Paolo Ferragina} {and}
  \bibinfo{person}{Giorgio Vinciguerra}.} \bibinfo{year}{2021}\natexlab{}.
\newblock \bibinfo{booktitle}{\emph{PGM-index: State-of-the-art learned data
  structure}}.
\newblock
\urldef\tempurl%
\url{https://github.com/gvinciguerra/PGM-index}
\showURL{%
\tempurl}


\bibitem[Galakatos et~al\mbox{.}(2019)]%
        {RN410}
\bibfield{author}{\bibinfo{person}{Alex Galakatos}, \bibinfo{person}{Michael
  Markovitch}, \bibinfo{person}{Carsten Binnig}, \bibinfo{person}{Rodrigo
  Fonseca}, {and} \bibinfo{person}{Tim Kraska}.}
  \bibinfo{year}{2019}\natexlab{}.
\newblock \bibinfo{title}{FITing-Tree: A Data-aware Index Structure}.
\newblock , \bibinfo{numpages}{1189–1206}~pages.
\newblock
\urldef\tempurl%
\url{https://doi.org/10.1145/3299869.3319860}
\showDOI{\tempurl}


\bibitem[Goldblum et~al\mbox{.}(2020)]%
        {RN447}
\bibfield{author}{\bibinfo{person}{Micah Goldblum}, \bibinfo{person}{Dimitris
  Tsipras}, \bibinfo{person}{Chulin Xie}, \bibinfo{person}{Xinyun Chen},
  \bibinfo{person}{Avi Schwarzschild}, \bibinfo{person}{Dawn Song},
  \bibinfo{person}{Aleksander Madry}, \bibinfo{person}{Bo Li}, {and}
  \bibinfo{person}{Tom Goldstein}.} \bibinfo{year}{2020}\natexlab{}.
\newblock \showarticletitle{Data Security for Machine Learning: Data Poisoning,
  Backdoor Attacks, and Defenses}.
\newblock \bibinfo{journal}{\emph{arXiv preprint arXiv:2012.10544}}
  (\bibinfo{year}{2020}).
\newblock


\bibitem[Hadian and Heinis(2019)]%
        {RN455}
\bibfield{author}{\bibinfo{person}{Ali Hadian} {and} \bibinfo{person}{Thomas
  Heinis}.} \bibinfo{year}{2019}\natexlab{}.
\newblock \showarticletitle{Interpolation-friendly B-trees: Bridging the Gap
  Between Algorithmic and Learned Indexes}.
\newblock  (\bibinfo{year}{2019}).
\newblock


\bibitem[Ho et~al\mbox{.}(2019)]%
        {RN405}
\bibfield{author}{\bibinfo{person}{Darryl Ho}, \bibinfo{person}{Jialin Ding},
  \bibinfo{person}{Sanchit Misra}, \bibinfo{person}{Nesime Tatbul},
  \bibinfo{person}{Vikram Nathan}, \bibinfo{person}{Vasimuddin Md}, {and}
  \bibinfo{person}{Tim Kraska}.} \bibinfo{year}{2019}\natexlab{}.
\newblock \showarticletitle{LISA: towards learned DNA sequence search}.
\newblock \bibinfo{journal}{\emph{arXiv preprint arXiv:1910.04728}}
  (\bibinfo{year}{2019}).
\newblock


\bibitem[Huang et~al\mbox{.}(2011)]%
        {RN701}
\bibfield{author}{\bibinfo{person}{Ling Huang}, \bibinfo{person}{Anthony~D
  Joseph}, \bibinfo{person}{Blaine Nelson}, \bibinfo{person}{Benjamin~IP
  Rubinstein}, {and} \bibinfo{person}{J~Doug Tygar}.}
  \bibinfo{year}{2011}\natexlab{}.
\newblock \showarticletitle{Adversarial machine learning}. In
  \bibinfo{booktitle}{\emph{Proceedings of the 4th ACM workshop on Security and
  artificial intelligence}}. \bibinfo{pages}{43--58}.
\newblock


\bibitem[Jagielski et~al\mbox{.}(2018)]%
        {RN461}
\bibfield{author}{\bibinfo{person}{Matthew Jagielski}, \bibinfo{person}{Alina
  Oprea}, \bibinfo{person}{Battista Biggio}, \bibinfo{person}{Chang Liu},
  \bibinfo{person}{Cristina Nita-Rotaru}, {and} \bibinfo{person}{Bo Li}.}
  \bibinfo{year}{2018}\natexlab{}.
\newblock \showarticletitle{Manipulating machine learning: Poisoning attacks
  and countermeasures for regression learning}. In
  \bibinfo{booktitle}{\emph{2018 IEEE Symposium on Security and Privacy (SP)}}.
  \bibinfo{publisher}{IEEE}, \bibinfo{pages}{19--35}.
\newblock
\showISBNx{1538643537}


\bibitem[Kipf et~al\mbox{.}(2020)]%
        {RN406}
\bibfield{author}{\bibinfo{person}{Andreas Kipf}, \bibinfo{person}{Ryan
  Marcus}, \bibinfo{person}{Alexander van Renen}, \bibinfo{person}{Mihail
  Stoian}, \bibinfo{person}{Alfons Kemper}, \bibinfo{person}{Tim Kraska}, {and}
  \bibinfo{person}{Thomas Neumann}.} \bibinfo{year}{2020}\natexlab{}.
\newblock \showarticletitle{RadixSpline: a single-pass learned index}. In
  \bibinfo{booktitle}{\emph{Proceedings of the Third International Workshop on
  Exploiting Artificial Intelligence Techniques for Data Management}}.
  \bibinfo{pages}{1--5}.
\newblock


\bibitem[Kornaropoulos et~al\mbox{.}(2020)]%
        {RN415}
\bibfield{author}{\bibinfo{person}{Evgenios~M Kornaropoulos},
  \bibinfo{person}{Silei Ren}, {and} \bibinfo{person}{Roberto Tamassia}.}
  \bibinfo{year}{2020}\natexlab{}.
\newblock \showarticletitle{The Price of Tailoring the Index to Your Data:
  Poisoning Attacks on Learned Index Structures}.
\newblock \bibinfo{journal}{\emph{arXiv preprint arXiv:2008.00297}}
  (\bibinfo{year}{2020}).
\newblock


\bibitem[Kraska et~al\mbox{.}(2018)]%
        {RN385}
\bibfield{author}{\bibinfo{person}{Tim Kraska}, \bibinfo{person}{Alex Beutel},
  \bibinfo{person}{Ed~H Chi}, \bibinfo{person}{Jeffrey Dean}, {and}
  \bibinfo{person}{Neoklis Polyzotis}.} \bibinfo{year}{2018}\natexlab{}.
\newblock \showarticletitle{The case for learned index structures}. In
  \bibinfo{booktitle}{\emph{Proceedings of the 2018 International Conference on
  Management of Data}}. \bibinfo{pages}{489--504}.
\newblock


\bibitem[Liu et~al\mbox{.}(2017)]%
        {RN459}
\bibfield{author}{\bibinfo{person}{Chang Liu}, \bibinfo{person}{Bo Li},
  \bibinfo{person}{Yevgeniy Vorobeychik}, {and} \bibinfo{person}{Alina Oprea}.}
  \bibinfo{year}{2017}\natexlab{}.
\newblock \showarticletitle{Robust linear regression against training data
  poisoning}. In \bibinfo{booktitle}{\emph{Proceedings of the 10th ACM Workshop
  on Artificial Intelligence and Security}}. \bibinfo{pages}{91--102}.
\newblock


\bibitem[Macke et~al\mbox{.}(2018)]%
        {RN457}
\bibfield{author}{\bibinfo{person}{Stephen Macke}, \bibinfo{person}{Alex
  Beutel}, \bibinfo{person}{Tim Kraska}, \bibinfo{person}{Maheswaran
  Sathiamoorthy}, \bibinfo{person}{Derek~Zhiyuan Cheng}, {and}
  \bibinfo{person}{EH Chi}.} \bibinfo{year}{2018}\natexlab{}.
\newblock \showarticletitle{Lifting the curse of multidimensional data with
  learned existence indexes}. In \bibinfo{booktitle}{\emph{Workshop on ML for
  Systems at NeurIPS}}.
\newblock


\bibitem[Marcus et~al\mbox{.}(2020)]%
        {RN386}
\bibfield{author}{\bibinfo{person}{Ryan Marcus}, \bibinfo{person}{Emily Zhang},
  {and} \bibinfo{person}{Tim Kraska}.} \bibinfo{year}{2020}\natexlab{}.
\newblock \showarticletitle{CDFShop: Exploring and Optimizing Learned Index
  Structures}. In \bibinfo{booktitle}{\emph{Proceedings of the 2020 ACM SIGMOD
  International Conference on Management of Data}}.
  \bibinfo{pages}{2789--2792}.
\newblock


\bibitem[Mei and Zhu(2015)]%
        {RN703}
\bibfield{author}{\bibinfo{person}{Shike Mei} {and} \bibinfo{person}{Xiaojin
  Zhu}.} \bibinfo{year}{2015}\natexlab{}.
\newblock \showarticletitle{Using machine teaching to identify optimal
  training-set attacks on machine learners}. In
  \bibinfo{booktitle}{\emph{Twenty-Ninth AAAI Conference on Artificial
  Intelligence}}.
\newblock


\bibitem[Microsoft(2021)]%
        {sourcecode-alex}
\bibfield{author}{\bibinfo{person}{Microsoft}.}
  \bibinfo{year}{2021}\natexlab{}.
\newblock \bibinfo{booktitle}{\emph{ALEX - A library for building an in-memory,
  Adaptive Learned indEX}}.
\newblock
\urldef\tempurl%
\url{https://github.com/microsoft/ALEX}
\showURL{%
\tempurl}


\bibitem[Mitzenmacher(2019)]%
        {RN429}
\bibfield{author}{\bibinfo{person}{Michael Mitzenmacher}.}
  \bibinfo{year}{2019}\natexlab{}.
\newblock \showarticletitle{A model for learned bloom filters, and optimizing
  by sandwiching}.
\newblock \bibinfo{journal}{\emph{arXiv preprint arXiv:1901.00902}}
  (\bibinfo{year}{2019}).
\newblock


\bibitem[Muñoz-González et~al\mbox{.}(2017)]%
        {RN705}
\bibfield{author}{\bibinfo{person}{Luis Muñoz-González},
  \bibinfo{person}{Battista Biggio}, \bibinfo{person}{Ambra Demontis},
  \bibinfo{person}{Andrea Paudice}, \bibinfo{person}{Vasin Wongrassamee},
  \bibinfo{person}{Emil~C Lupu}, {and} \bibinfo{person}{Fabio Roli}.}
  \bibinfo{year}{2017}\natexlab{}.
\newblock \showarticletitle{Towards poisoning of deep learning algorithms with
  back-gradient optimization}. In \bibinfo{booktitle}{\emph{Proceedings of the
  10th ACM Workshop on Artificial Intelligence and Security}}.
  \bibinfo{pages}{27--38}.
\newblock


\bibitem[Müller et~al\mbox{.}(2020)]%
        {RN715}
\bibfield{author}{\bibinfo{person}{Nicolas Müller}, \bibinfo{person}{Daniel
  Kowatsch}, {and} \bibinfo{person}{Konstantin Böttinger}.}
  \bibinfo{year}{2020}\natexlab{}.
\newblock \showarticletitle{Data Poisoning Attacks on Regression Learning and
  Corresponding Defenses}. In \bibinfo{booktitle}{\emph{2020 IEEE 25th Pacific
  Rim International Symposium on Dependable Computing (PRDC)}}.
  \bibinfo{publisher}{IEEE}, \bibinfo{pages}{80--89}.
\newblock
\showISBNx{1728180031}


\bibitem[Nathan et~al\mbox{.}(2020)]%
        {RN407}
\bibfield{author}{\bibinfo{person}{Vikram Nathan}, \bibinfo{person}{Jialin
  Ding}, \bibinfo{person}{Mohammad Alizadeh}, {and} \bibinfo{person}{Tim
  Kraska}.} \bibinfo{year}{2020}\natexlab{}.
\newblock \showarticletitle{Learning multi-dimensional indexes}. In
  \bibinfo{booktitle}{\emph{Proceedings of the 2020 ACM SIGMOD International
  Conference on Management of Data}}. \bibinfo{pages}{985--1000}.
\newblock


\bibitem[Setiawan et~al\mbox{.}(2020)]%
        {RN408}
\bibfield{author}{\bibinfo{person}{Naufal~Fikri Setiawan},
  \bibinfo{person}{Benjamin~IP Rubinstein}, {and} \bibinfo{person}{Renata
  Borovica-Gajic}.} \bibinfo{year}{2020}\natexlab{}.
\newblock \showarticletitle{Function interpolation for learned index
  structures}. In \bibinfo{booktitle}{\emph{Australasian Database Conference}}.
  \bibinfo{publisher}{Springer}, \bibinfo{pages}{68--80}.
\newblock


\bibitem[Tang et~al\mbox{.}(2019a)]%
        {RN453}
\bibfield{author}{\bibinfo{person}{Chuzhe Tang}, \bibinfo{person}{Zhiyuan
  Dong}, \bibinfo{person}{Minjie Wang}, \bibinfo{person}{Zhaoguo Wang}, {and}
  \bibinfo{person}{Haibo Chen}.} \bibinfo{year}{2019}\natexlab{a}.
\newblock \showarticletitle{Learned indexes for dynamic workloads}.
\newblock \bibinfo{journal}{\emph{arXiv preprint arXiv:1902.00655}}
  (\bibinfo{year}{2019}).
\newblock


\bibitem[Tang et~al\mbox{.}(2019b)]%
        {RN725}
\bibfield{author}{\bibinfo{person}{Sanli Tang}, \bibinfo{person}{Xiaolin
  Huang}, \bibinfo{person}{Mingjian Chen}, \bibinfo{person}{Chengjin Sun},
  {and} \bibinfo{person}{Jie Yang}.} \bibinfo{year}{2019}\natexlab{b}.
\newblock \showarticletitle{Adversarial attack type i: Cheat classifiers by
  significant changes}.
\newblock \bibinfo{journal}{\emph{IEEE transactions on pattern analysis and
  machine intelligence}} (\bibinfo{year}{2019}).
\newblock
\showISSN{0162-8828}


\bibitem[Xiao et~al\mbox{.}(2015)]%
        {RN706}
\bibfield{author}{\bibinfo{person}{Huang Xiao}, \bibinfo{person}{Battista
  Biggio}, \bibinfo{person}{Gavin Brown}, \bibinfo{person}{Giorgio Fumera},
  \bibinfo{person}{Claudia Eckert}, {and} \bibinfo{person}{Fabio Roli}.}
  \bibinfo{year}{2015}\natexlab{}.
\newblock \showarticletitle{Is feature selection secure against training data
  poisoning?}. In \bibinfo{booktitle}{\emph{international conference on machine
  learning}}. \bibinfo{publisher}{PMLR}, \bibinfo{pages}{1689--1698}.
\newblock


\bibitem[Zhang et~al\mbox{.}(2021)]%
        {RN721}
\bibfield{author}{\bibinfo{person}{Zhou Zhang}, \bibinfo{person}{Pei-Quan Jin},
  \bibinfo{person}{Xiao-Liang Wang}, \bibinfo{person}{Yan-Qi Lv},
  \bibinfo{person}{Shou-Hong Wan}, {and} \bibinfo{person}{Xi-Ke Xie}.}
  \bibinfo{year}{2021}\natexlab{}.
\newblock \showarticletitle{COLIN: A Cache-Conscious Dynamic Learned Index with
  High Read/Write Performance}.
\newblock \bibinfo{journal}{\emph{Journal of Computer Science and Technology}}
  \bibinfo{volume}{36}, \bibinfo{number}{4} (\bibinfo{year}{2021}),
  \bibinfo{pages}{721--740}.
\newblock
\showISSN{1860-4749}


\end{thebibliography}

\begin{acronym}
	\acro{ALEX}[ALEX]{Adaptive Learned indEX}
	\acro{CDF}[CDF]{Cumulative Distribution Function}
	\acro{GA}[GA]{Gapped Array}
	\acro{LID}[LID]{Local Intrinsic Dimensionality}
	\acro{LIS}[LIS]{Learned Index Structure}
	\acro{LS}[LS]{Linear Spline}
	\acro{MSE}[MSE]{Mean Squared Error}
	\acro{ML}[ML]{Machine Learning}
	\acro{PGM}[PGM]{Piecewise Geometric Model}
	\acro{PLA}[PLA]{Piecewise Linear Approximation}
	\acro{RMI}[RMI]{Recursive Model Index}
\end{acronym}

\end{document}